\documentclass[12pt,preprint]{aastex}
\usepackage[dvips]{color}
\usepackage{graphicx}

\def\msol{\hbox{\kern 0.20em $M_\odot$}}
\def\lsol{\hbox{\kern 0.20em $L_\odot$}}
\def\rsol{\hbox{\kern 0.20em $R_\odot$}}
\def\sr{\hbox{\kern 0.20em sr}}
\def\srmu{\hbox{\kern 0.20em sr$^{-1}$}}

\def\g{\hbox{\kern 0.20em g}}
\def\gmu{\hbox{\kern 0.20em g$^{-1}$}}
\def\kg{\hbox{\kern 0.20em kg}}
\def\pc{\hbox{\kern 0.20em pc}}

\def\mum{\hbox{\kern 0.20em $\mu$m}}
\def\mumd{\hbox{\kern 0.20em $\mu$m$^{-2}$}}
\def\cm{\hbox{\kern 0.20em cm}}
\def\m{\hbox{\kern 0.20em m}}
\def\km{\hbox{\kern 0.20em km}}
\def\nm{\hbox{\kern 0.20em nm}}

\def\s{\hbox{\kern 0.20em s}}
\def\h{\hbox{\kern 0.20em h}}
\def\sec{\hbox{\kern 0.20em sec}}
\def\min{\hbox {\kern 0.20em min}}
\def\smu{\hbox{\kern 0.20em s$^{-1}$}}
\def\smd{\hbox{\kern 0.20em s$^{-2}$}}
\def\an{\hbox{\kern 0.20em an}}
\def\anmu{\hbox{\kern 0.20em an$^{-1}$}}
\def\deg{\hbox{\kern 0.20em $^{\rm o}$}}
\def\yr{\hbox{\kern 0.20em yr}}
\def\yrmu{\hbox{\kern 0.20em yr$^{-1}$}}
\def\Myr{\hbox{\kern 0.20em Myr}}
\def\Mymu{\hbox{\kern 0.20em Myr$^{-1}$}}
\def\K{\hbox{\kern 0.20em K}}
\def\pcmu{\hbox{\kern 0.20em pc$^{-1}$}}
\def\pcmd{\hbox{\kern 0.20em pc$^{-2}$}}
\def\pcmt{\hbox{\kern 0.20em pc$^{-3}$}}
\def\kms{\hbox{\kern 0.20em km\kern 0.20em s$^{-1}$}}
\def\kmpd{\hbox{\kern 0.20em km$^{2}$}}
\def\kpc{\hbox{\kern 0.20em kpc}}
\def\cms{\hbox{\kern 0.20em cm\kern 0.20em s$^{-1}$}}
\def\erg{\hbox{\kern 0.20em erg}}
\def\ergs{\hbox{\kern 0.20em erg}}
\def\cmpd{\hbox{\kern 0.20em cm$^2$}}
\def\cmmd{\hbox{\kern 0.20em cm$^{-2}$}}
\def\cmms{\hbox{\kern 0.20em cm$^{-6}$}}
\def\cmpt{\hbox{\kern 0.20em cm$^3$}}
\def\cmmt{\hbox{\kern 0.20em cm$^{-3}$}}
\def\mpd{\hbox{\kern 0.20em m$^2$}}
\def\mmd{\hbox{\kern 0.20em m$^{-2}$}}
\def\mpt{\hbox{\kern 0.20em m$^3$}}
\def\mmt{\hbox{\kern 0.20em m$^{-3}$}}
\def\mujy{\hbox{\kern 0.20em $\mu$Jy}}
\def\mjy{\hbox{\kern 0.20em mJy}}
\def\Mj{\hbox{\kern 0.20em MJy}}
\def\jy{\hbox{\kern 0.20em Jy}}
\def\ghz{\hbox{\kern 0.20em GHz}}
\def\srmd{\hbox{\kern 0.20em sr$^{-1}$}}
\def \kms{km~$\rm{s}^{-1}$}

\def \mum{$\mu$m}
\def\hii{\relax \ifmmode {\rm H\,{\sc ii}}\else H\,{\sc ii}\fi}

\def\G{\hbox{\kern 0.20em G}}

\def\h13cop{\hbox{H$^{13}$CO$^{+}$}}

\def\S+{\hbox{S{\small II}}}

\slugcomment{The Evolving ISM in the Milky Way \& Nearby Galaxies}

\shorttitle{Tracing the [FeII]/[NeII] ratio within star forming dwarf galaxies: a \emph{Spitzer} IRS archival study.}

\shortauthors{O'Halloran et al.}

\begin{document}

\newcommand{\jfourteen}{\hbox{$J=14\rightarrow 13$}}
 \title{Tracing the [FeII]/[NeII] ratio and its relationship with other ISM indicators within star forming dwarf galaxies: a \emph{Spitzer} IRS archival study.}

\author{B. O'Halloran\altaffilmark{1},  S. C. Madden\altaffilmark{2},  N. P. Abel\altaffilmark{3}}

\email{boh@physics.gmu.edu}

\altaffiltext{1}{Dept. of Physics \& Astronomy, George Mason University, Fairfax, VA 22030, USA}

\altaffiltext{2}{Service d'Astrophysique, CEA, Saclay, Orme des Merisiers 91191, Gif-sur-Yvette, France}

\altaffiltext{3}{Department of Physics, University of Cincinnati, Cincinnati, OH 45221, USA}

\begin{abstract}

Archival \emph{Spitzer} observations of 41 starburst galaxies that span a wide range in metallicity reveal for the first time a correlation between
the [FeII]/[NeII] 26.0/12.8 $\mu$m ratio and the electron gas density as traced by the 18.7/33.4 $\mu$m [SIII] ratio, with the [FeII]/[NeII] ratio
decreasing with increasing gas density. We also find a strong correlation between the gas density and the PAH peak to continuum strength. Using shock
and photoionization models, we see the driver of the observed [FeII]/[NeII] ratios is metallicity. The majority of [FeII] emission in low metallicity
galaxies may be shock-derived, whilst at high metallicity, the [FeII] emission may be instead dominated by contributions from \hii\ and in particular
from dense PDR regions. However, the observed [FeII]/[NeII] ratios may instead be following a metallicity-abundance relationship, with iron being
less depleted onto grains in low metallicity galaxies - a result that would have profound implications for the use of iron emission lines as
unambiguous tracers of shocks.

\end{abstract}


\lefthead{O'Halloran et al.}

\righthead{Tracing the [FeII]/[NeII] ratio within star forming dwarf galaxies: a \emph{Spitzer} IRS archival study.}

\section{Introduction}

The presence of massive stars within starbursts undoubtedly plays a huge role in determining the physical conditions within the local interstellar
medium (ISM). High-mass (M$_{init}$ $\geq$ 8M$_{\odot}$) stars formed within typical starbursts drastically affect the dynamics of the surrounding
ISM, through not only the release of ionizing photons which destroy molecular material, but also via supernovae (SNe) which provide thermal and
kinetic energy input into the ISM. The effects of photoionization by high-mass stars on the observed dearth of PAHs have recently been investigated
(Madden et al. (2006); Wu et al. (2006)), while in O'Halloran et al. (2006) (hereafter Paper I), we examined a sample of 18 galaxies of varying
metallicity (from 1/50th to super solar) with high star formation rates in order to determine whether supernova-driven shocks do indeed play a role
in the PAH deficit in low metallicity environments. If we consider the ratio of the 26 $\mu$m [FeII] line and the 12.8 $\mu$m [NeII] line as a tracer
of the strength of supernova shocks, we found a strong anti-correlation suggesting that strong supernova-driven shocks are indeed present within low
metallicity galaxies. Furthermore, the PAH deficit within these objects may indeed be linked to the presence and strength of these shocks. However,
it has not as yet been conclusively proved that shocks are the dominant process behind the PAH deficit. It would therefore be advisable to further
explore the nature of the ISM within such star-forming environments using additional mid-IR probes in order to expand upon our understanding of the
[FeII]/[NeII] ratio, and by extension, its relationship with key ISM indicators such as PAH strength and metallicity.

\section{Sample and Data Reduction}

To accomplish this, we have expanded our sample from the 18 objects presented in Paper I by including archival IRS observations of 23 additional
objects, bringing the sample total to 41. These galaxies range in metallicity from extremely low (such as I Zw 18, with $ \it Z/Z_{\odot}$=  1/50) to
super-solar metallicity ($\geq 1~\it Z_{\odot}$) galaxies such as NGC 7714.  None of the galaxies in our sample are known to harbour AGNs. We
extracted low  and high resolution archival spectral  data from the Short-Low  (SL) (5.2 -  14.5  $\mu$m), Short-High (SH) (9.9 - 19.6 $\mu$m) and
Long-High (LH) (18.7 -  37.2 $\mu$m) modules  of the {\it Spitzer} Infrared Spectrograph (IRS).   The datasets were derived from a number of {\it
Spitzer}  Legacy, GO and GTO programs released to the  {\it Spitzer} Data  Archive, and consisted of  either spectral  mapping  or staring
observations. We obtained fluxes for the nuclear positions only from the mapping observations. All the staring observations were centered on the
galaxy's nucleus.

\section{Results and Discussion}

In Fig 1, (top left) we plot [FeII]/[NeII] vs the 6.2 $\mu$m PAH peak to continuum ratio to check if the relationship between the [FeII]/[NeII] ratio
and the PAH peak to continuum strength first noted in Paper I. Employing a Spearman rank correlation analysis to assess the statistical significance
of this trend yields a correlation coefficient of r$_{s}$ of -0.705 and P$_{s}$ of 2.69 x 10$^{-6}$, confirming the significant anti-correlation
between PAH strength and the [FeII]/[NeII] ratio as seen in Paper I. We plot the [FeII]/[NeII] ratio versus the metallicity in Fig.1 (top right), and
we again see a strong anti-correlation (r$_{s}$ of -0.777 and P$_{s}$ of 9.36 x 10$^{-7}$) between [FeII]/[NeII] and metallicity as seen in paper I.

Based upon what we have already seen from Paper I and from the figures, if the [FeII]/[NeII] ratio is truly indicative of the strength of
supernova-driven shocks within the extent of the high resolution slits, one would expect the passage of such shocks to affect conditions within the
local ISM in quite a substantive manner. Shocks, in addition to removing dust and PAH from the ISM, should also be adept in removing gas - one would
therefore expect the propagation of intense SNe-driven shocks into the ISM of star forming regions to greatly impact on the density of the gas. In
order to determine how the gas density within these nuclear star forming regions corresponds with the strength of the supernova-driven shocks, we
require a reliable mid-IR tracer to probe the gas density. The [SIII] 18.7/33.4 $\mu$m line ratio provides such a reliable mid-IR tracer of the gas
density, as it is ideal for probing gas (with high critical densities) in the regions surrounding starbursts, especially as the [SIII] lines are
starburst dominated. This ratio is sensitive to changes in the density for the 50 $\leq$ n$_{e}$ $\leq$ 10$^{4}$ cm$^{-3}$, but is insensitive to
changes in temperature. We also plot the logarithm of the ratio of the [SIII] fluxes versus the [FeII]/[NeII] flux ratio (bottom left). There is a
strikingly strong trend between the two line ratios, with high [FeII]/[NeII] values corresponding to lower [SIII] ratios. Using the Spearman rank
correlation analysis, we get (r$_{s}$) of -0.891 between the [SIII] ratio and the [FeII]/[NeII] with a probability of chance correlation (P$_{s}$) of
7.43 x 10$^{-8}$, indicating a significant anti-correlation. Interestingly, we see a general decrease in the metallicity of the object with lower
[SIII] ratios, corresponding to higher [FeII]/[NeII].

We again plot the logarithm of the ratio of the [SIII] fluxes, but this time against the 6.2 micron PAH peak/continuum ratio (bottom right Fig. 1).
We see a strong trend, but this time with the PAH strength increasing with log [SIII] - the PAH strength is increasing with higher gas density. Again
running a Spearman rank correlation analysis, we get r$_{s}$ of 0.750 and P$_{s}$ of 9.64 x 10$^{-7}$, confirming a significant correlation between
log [SIII] and PAH strength. Breaking down the points in metallicity bins, we again see that a general trend exists with PAH strength increasing with
metallicity and increasing values of log [SIII].

Ruling out extinction and aperture effects as drivers for the observed correlations, we can be confident that the correlations seen in Fig. 1 are
indeed true physical relationships between the strength of the [FeII]/[NeII] ratio and a number of ISM indicators, with the primary culprit for the
observed relationships being the passage of SNe-driven shocks. However, while we have focussed up to now with a supernova-derived shock origin for
the behaviour of the [FeII]/[NeII] ratio, it may not be the only explanation. For example, Izotov et al. (2006) note sign of strong depletion of iron
onto dust grains, and gradual destruction of those grains on a time scale of a few Myr, based on a survey of metal-poor galaxies from the 3rd release
of the SDSS. Such a process could undoubtedly drastically affect the nature of Fe emission, and by extension the behaviour of the [FeII]/[NeII]
ratio, within our sample - we may instead be probing an abundance-driven relationship and the other observed relationships presented here would be
purely coincidental to the [FeII]/[NeII] ratio. In order to explore if this indeed is the true cause or if shocks alone can explain the observed
relationships, we used shock and standard \hii\ - PDR models in an effort to model the observed [FeII]/[NeII] relationship for a wide variety of
environments. We used both the \emph{MAPPINGS III} photoionization/shock code and the Cloudy photoionization code in order to determine the relative
proportions to the [FeII] emission from shocks and \hii/PDRs, and by extension, the driving process behind the behaviour of the observed
[FeII]/[NeII] line ratio. We found that the relative \hii\ contribution dominates over the PDR contribution at low densities (and metallicities),
with the PDR contribution becoming more dominant as one moves to higher densities ($\sim$ 10$^{3}$ cm$^{-3}$) and metallicities. For very dense \hii\
regions where the PDRs are irradiated by intense FUV, and thus PDR-derived [FeII] emission is dominant.  Correspondingly at lower densities
($\sim$10$^{2}$ cm$^{-3}$), the \hii\ regions are larger, the PDRs lie further from the stars, and the resultant lowered FUV fluxes and densities do
not excite [FeII] in the PDR, and thus the \hii\ region dominates the \hii/PDR contribution to [FeII] production for such environments. From the $\it
MAPPINGS$ and $\it Cloudy$ simulations and the comparisons with the IRS and optical data, it would seem that metallicity is the primary driver for
the behaviour of the [FeII]/[NeII] ratio, as metallicity determines the origin (shocks vs \hii\ and/or PDRs) and strength of the [FeII] emission. We
can reproduce the observed [FeII]/[NeII] dependence on \emph{Z} if we assume the Izotov variation of \emph{Fe/Ne} with \emph{Z}. However, we do note
that the \hii\ and PDR contribution to the [FeII] emission is surpassed by the contribution from shocks at the lowest metallicities and densities,
based on comparison with the \emph{MAPPINGS} output and the observed data - perhaps only $\sim$10\% of the [FeII] emission comes from \hii\ and PDRs
with that contribution rising to $\sim$80-90\% as one moves to higher metallicity. As we move to roughly solar metallicity, it would seem likely that
the dominant mechanism for production of [FeII] (and thus the behaviour of the [FeII]/[NeII] ratio at high metallicities) is from \hii\ and PDRs in
combination rather than from SNe shocks.

\begin{figure}
\centerline{
\includegraphics*{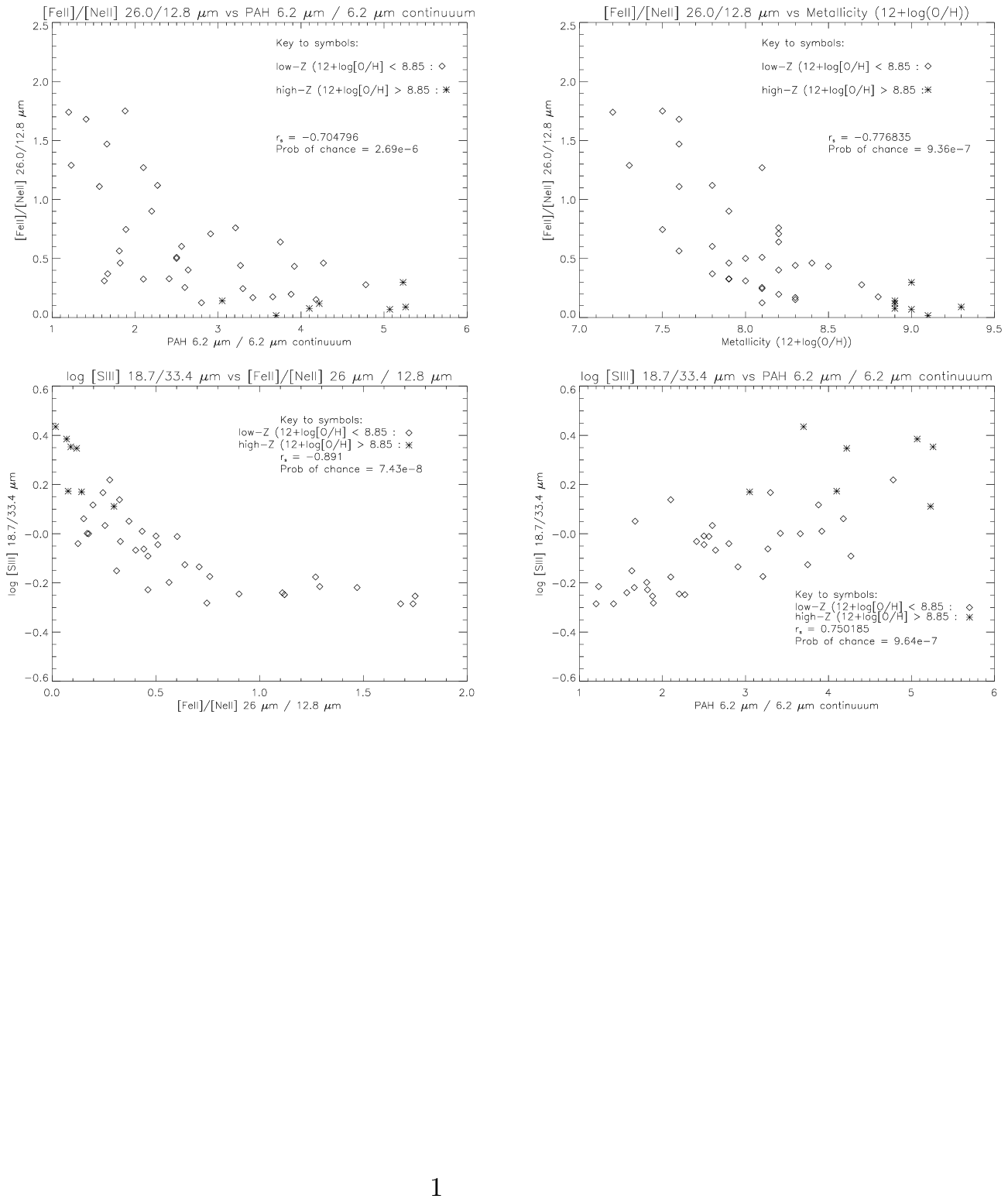}
} \caption{\label{fig:1} (Top left)Plot of the [FeII]/[NeII] ratio versus the PAH 6.2 micron peak to continuum ratio for the extended sample. (Top
right)Plot of the [FeII]/[NeII] ratio versus the metallicity for the extended sample. (Bottom left) Plot of the 18.7/33.4 $\mu$m [SIII] ratio as a
function of the [FeII]/[NeII] ratio for the extended sample. (Bottom right) Plot of the 18.7/33.4 $\mu$m [SIII] ratio as a function of the 6.2 $\mu$m
PAH peak to continuum strength for the extended sample. }
\end{figure}

\section{Summary}

Using archival \emph{Spitzer} observations of 41 starburst galaxies that span a wide range in metallicity, we found a correlation between the ratio
of emission line fluxes of [FeII] at 26 $\mu$m and [NeII] at 12.8 $\mu$m and the electron gas density as traced by the 18.7/33.4 $\mu$m [SIII] ratio,
with the [FeII]/[NeII] flux ratio decreasing with increasing gas density. We also find a strong correlation between the gas density and the PAH peak
to continuum strength. The correlation of the [FeII]/[NeII] ratio and the PAH peak to continuum strength found in paper I was confirmed for a larger
sample. Using shock and photoionization models, we see that metallicity is the primary driver for the observed behaviour of the [FeII]/[NeII] ratio.
It may very well be that the majority of [FeII] emission at low metallicity may be shock-derived, whilst at high metallicity, the [FeII] emission is
dominated by contributions from \hii\ and in particular, from PDR regions, and that at higher metallicity shocks may not play as significant a role
in removing gas, PAH and dust from the ISM, unlike in low metallicity systems. However, the observed [FeII]/[NeII] emission may instead be following
a metallicity-abundance relationship, with the iron being less depleted in low metallicity galaxies, a result that would have profound implications
for the use of Fe emission lines as unambiguous tracers of shocks.
\acknowledgments

We are grateful to S. Satyapal for her comments and guidance. BO'H gratefully acknowledges financial support from NASA grant NAG5-11432. NPA
acknowledges National Science Foundation support under Grant No. 0094050 , 0607497 to The University of Cincinnati. This work is based on archival
data obtained with the \emph{Spitzer} Space Telescope, which is operated by the Jet Propulsion Laboratory, California Institute of Technology under a
contract with NASA.


\begin{references}{}

\reference{galliano08} Galliano F., Dwek E. \& Chanial P., 2008, \apj, 672, 214

\reference{izotov06} Izotov Y.I., Stasinska G., Meynet G. et al., 2006, \aap, 448, 955

\reference{oha06} O'Halloran B., Satyapal S. \& Dudik R.P., 2006, \apj, 641, 795

\reference{wu06} Wu Y., Charmandaris V., Hao L. et al., 2006, \apj, 639, 157

\end{references}
\end{document}